\documentclass{IEEEcsmag}

\usepackage[colorlinks,urlcolor=blue,linkcolor=blue,citecolor=blue]{hyperref}

\usepackage{upmath}

\jvol{XX}
\jnum{XX}
\paper{8}
\jmonth{January/February}
\jname{IEEE Computer Graphics and Applications}
\pubyear{2024}

\newcommand{\edits}[1]{\textcolor{black}{{{#1}}}}

\begin{document}

\sptitle{DEPARTMENT: VISUALIZATION VIEWPOINTS}

\title{Using Counterfactuals to Improve Causal Inferences from Visualizations}

\author{David Borland}
\affil{RENCI, University of North Carolina at Chapel Hill, Chapel Hill, NC, 27599, USA}

\author{Arran Zeyu Wang}
\affil{University of North Carolina at Chapel Hill, Chapel Hill, NC, 27599, USA}

\author{David Gotz}
\affil{University of North Carolina at Chapel Hill, Chapel Hill, NC, 27599, USA}

\markboth{VISUALIZATION VIEWPOINTS}{VISUALIZATION VIEWPOINTS}

\begin{abstract}
Traditional approaches to data visualization have often focused on comparing different subsets of data, and this is reflected in the many techniques developed and evaluated over the years for visual comparison. Similarly, common workflows for exploratory visualization are built upon the idea of users interactively applying various filter and grouping mechanisms in search of new insights. This paradigm has proven effective at helping users identify correlations between variables that can inform thinking and decision-making. However, recent studies show that consumers of visualizations often draw causal conclusions even when not supported by the data. Motivated by these observations, this article highlights recent advances from a growing community of researchers exploring methods that aim to directly support visual causal inference. However, many of these approaches have their own limitations which limit their use in many real-world scenarios. This article therefore also outlines a set of key open challenges and corresponding priorities for new research to advance the state of the art in visual causal inference.

\end{abstract}

\maketitle

\begin{figure*}[t]
\centerline{\includegraphics[width=38pc]{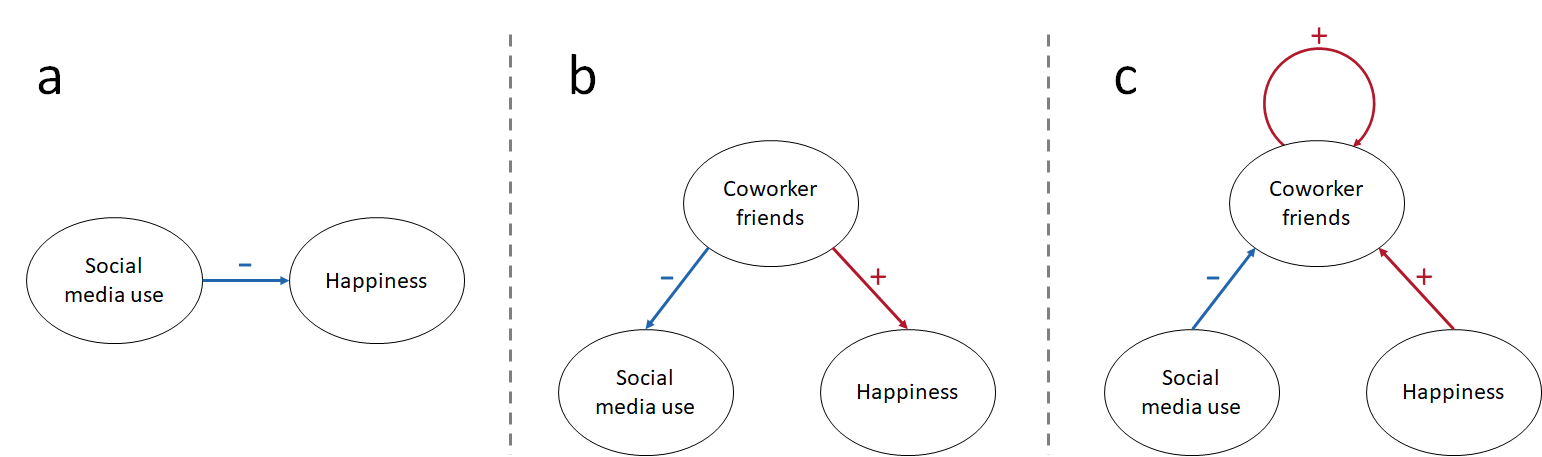}}
\caption{Example causal graphs: (a) a simple causal graph in which social media use decreases happiness, (b) a graph with a \emph{confounder}, in which the apparent causal relationship between social media use and happiness is in fact caused by a third factor, number of coworker friends, that has a causal effect on both, and (c) a graph with a \emph{collider}, in which the apparent causal relationship between social media use and happiness is due to both having a causal effect on this third factor. In this case, the collider also exhibits a cycle.
}
\label{causal_diagrams_fig}
\end{figure*}

\chapteri{D}ata visualization has become a ubiquitous tool for data exploration and communication for a wide range of audiences. Interactive visualizations now can be found in everything from GUI-based visual analysis tools for everyday office workers (e.g., Tableau), to websites produced by major media companies (e.g., FiveThirtyEight or the New York Times), and in notebook computing environments used by data scientists (e.g., \edits{RStudio, JupyterLab, and Observable}). Visualizations have also become key components in tools ranging from consumer-focused mobile apps to manage health and wellbeing (e.g., Apple's Health app) to expert-focused information systems such as electronic health record software used at most major hospitals (e.g., Epic). Visualizations have been used to help engineers diagnose and improve machine learning models, to help factories optimize manufacturing processes, and to to share news with the general public about the spread and risk of disease during the recent COVID-19 pandemic.  

In each of these applications, visualizations are leveraged to communicate data to a human consumer who then interprets what they see and draws conclusions based on the data. Often, these conclusions then motivate action. For instance, an analyst developing deep learning models might choose to adjust a parameter of their model after seeing a strange pattern in a visualization of model outputs. A manufacturing expert might decide to adjust a machine's settings to optimize production after visually analyzing data from sensors placed along an assembly line. A traveler may decide it's safer to travel on vacation to one city vs. another after viewing infection and hospitalization data on a COVID-19 dashboard.  

In each of these cases, users are making inferences about relationships between the variables displayed in a visualization and using those inferences to guide their decision-making process. The correctness of a user's decisions, therefore, is clearly contingent upon the correctness of the inferences that they make about the meaning of their visualized data.  
Along these lines, recent studies~\cite{kale_causal_2021,kaul_improving_2021,hullman_designing_2021} have shown that people often interpret visualized patterns as indicators of causal relationships between visualized variables. \emph{Critically, this inference of causality occurs even when such conclusions are not supported by the data or visualization}. 

One aspect of this problem is that users often draw conclusions that do not properly account for sample size \cite{kale_causal_2021,hullman_designing_2021,kim_bayesian_2019}. This challenge can be partly mitigated by enhancing visual representations of key information such as sample size or confidence intervals.

More difficult to solve, however, is a more fundamental issue: visualizations are generally designed to communicate correlation and not causation. More specifically, there could be mismatches between (1) a human user's tendency to draw causal conclusions, and (2) the design of a typical visualization which provides simplified overviews and/or narrowly filtered views of complex multi-variate datasets \cite{shneiderman_eyes_1996}. These types of visualizations generally fail to communicate the many interactions that can exist between explanatory variables, and---more importantly---fail to help users understand the effects of these interactions on patterns that are in fact visualized.  The result is that visualizations can dramatically mislead users into drawing erroneous causal inferences. 

For example, imagine two groups of individuals where group A is active on social media and group B is not. Presented with a chart showing that group A is overall unhappier, one might infer that social media activity determines individual levels of happiness. This relationship is depicted in \autoref{causal_diagrams_fig} (a). The absence of visual cues about the effect of other attributes of the two groups which may actually be causing the difference in happiness (e.g., that group A may also have closer friendships at work and in their local neighborhood) can lead to users making unsupported assumptions about social media as a causal factor in a person's happiness. Some possible causal relationships taking friendships at work into account are depicted in \autoref{causal_diagrams_fig} (b) and (c).

\edits{\emph{Counterfactual reasoning} is a central pillar in casual analysis that has been developed to assist in thinking about these types of scenarios~\cite{pearl_causal_2016,lewis_counterfactuals_2013,morgan_counterfactuals_2015}.  \emph{Counterfactuals} are hypothesized scenarios that enable us to examine potential outcomes across different scenarios in which only a particular factor is altered. Isolating this factor enables improved reasoning about and understanding of causal relationships involving this factor, e.g. whether or not an individual has coworker friends in the example above, and can be helpful to better understand causal relationships. Previous work has shown that counterfactuals can help improve causal inferences from visualizations~\cite{kaul_improving_2021}.}

In this article, we provide an overview of recent research examining causal inference from visualizations as well as key areas of related work. We then argue for research exploring new approaches that help mitigate the risks of incorrect causal inferences during exploratory analysis and data-driven decision-making. More specifically, we argue for research exploring a new approach to visualization which leverages the concept of counterfactual reasoning as a means to help users draw more robust and generalizable inferences from modern data visualizations.

\section{\MakeUppercase{Perceiving Causality}}

The expression ``correlation does not imply causation'' is an oft-repeated phrase warning against drawing causal conclusions based only on an observation of correlation between variables. As the statement suggests, there are many reasons correlations can appear in data that do not provide direct evidence of a causal relationship. Yet despite this warning, previous research has shown repeatedly that people do in fact tend to improperly assign causal meaning to correlations they observe when using data visualization.

This effect was perhaps most directly described by Xiong et al.~\cite{xiong_illusion_2020} who studied the illusory effect of visualizations to imply causal relationships across a number of different types of classical statistical charts (e.g., bar charts, line graphs, and scatter plots). They reported illusions of causality across all chart types, with users confusing correlation with causality.

The results from Xiong et al.'s study showed that the magnitude of the illusion of causality differed between different chart types. More specifically, the results suggested that the magnitude of the causal illusion was influenced both by the type of visual encoding employed in a chart as well as the level of aggregation. In general, the authors found that higher levels of aggregation tend to increase the implication of a causal relationship. Similarly, they found that line and dot encodings implied higher levels of causality than bar-based encodings. These findings provide some insights into how design choices influence perceived causality, but it remains that the illusion was present across all chart types and therefore cannot be easily ameliorated just by avoiding certain types of charts.

Another recent study conducted by Kale et al.~\cite{kale_causal_2021} provided further insights into perceived causality from visualizations \edits{by leveraging mathematical psychology~\cite{griffiths2005structure} and a causal support model}.
\edits{The study design adopted by Kale et al. included a ground truth for the level of causal support behind the data used in the study. This design allowed, in contrast to other studies (e.g.,~\cite{xiong_illusion_2020,kaul_improving_2021}), a comparison between that ground truth and a users' perceptions.
Their study shows variances by participants when evaluating Directed Acyclic Graphs (DAGs) that either incorporate or omit a causal linkage between two variables. Users' assessments frequently deviated from the probabilities assigned to each causal explanation, with instances of both overestimation and underestimation of the likelihood of causal relationships. These findings further highlight the challenge of accurately gauging the extent of evidential support that a specific dataset provides for a given causal explanation.}

Adding to this growing body of data, we published results from our own study~\cite{kaul_improving_2021} which also gathered empirical evidence that users tend to assign causal effects to variables used to group data within simple bar charts and line graphs. Interestingly, our study also hinted at some approaches to partially mitigate the effect which we'll discuss in more detail later in this article.

In the same year, Hullman and Gelman~\cite{hullman_designing_2021} argued for grounding the design of interactive exploratory visualization tools within formal theories of graphical inference. Their argument recognized the importance of accounting for how people draw inferences when they consume visualizations of data. Without this grounding, visual analytics tools would be at greater risk of inadvertently leading users to invalid inferences from their data.

Together, these studies provide compelling evidence that users of visualizations tend to assign (often incorrectly and without valid statistical support) causal explanations to visualizations of correlated data. This tendency to perceive causal effect replicates across a variety of tasks using different widely-used chart types, and puts users of visualizations at risk of drawing invalid causal inferences from their data even when current best practices are followed for constructing the visualizations.

\section{\MakeUppercase{Visual Causal Inference}}

As described in the prior section, users tend to infer the existence of causal relationships when interpreting even basic statistical charts based on correlations. These causal inferences are often incorrect and not supported by the data or the corresponding visualization designs. 

However, it is also true that causal inference is in fact a critical requirement in many use cases. For this reason, a number of efforts have been made to build visual analytics tools that directly support visual causal inference by adopting workflows analogous to classic statistical approaches to causal inference.

In statistics, a causal structure model is a mathematical representation that is applied to describe the causal relationships between a set of variables.
These relationships are typically structured into node-link causal diagrams, where nodes are variables and links indicate the possible causal relationships \autoref{causal_diagrams_fig}.
The most common representation is the Directed Acyclic Graph (DAG), which is restricted to directed edges and no cycles.
It assumes that each variable (node) is influenced by its direct parent nodes in the graph and that there are no hidden confounders or feedback loops. Alternatives relax the acyclic constraint, adopting a Directed Causal Graph (DCG) structure that permits cycles which can represent more complex causal relationships. 

Typical visual analytics systems that take causality into account employ DCGs as their underlying model of causality. These may exist internally to generate causal inference statistics which are then visualized, or the DCGs may be visualized directly via node-link diagrams. 
Existing visual analytics systems that employ causal graphs for user display and interaction include \cite{jin_visual_2021, xie_visual_2021, hoque_outcome_2022, wang_visual_2016}. 

\edits{The gold standard for exploring causal effects is the randomized control trial, in which hypothesized causal structures are typically defined \emph{a priori} based on theoretical understanding or prior evidence. Such studies are common in medical research~\cite{armitage2008statistical}. Alternatively, statistical mining methods are commonly applied to find the structure of causal models from datasets (e.g., \cite{jin_visual_2021} computed Hawkes processes~\cite{xu2016learning} while~\cite{xie_visual_2021} employed the F-GES~\cite{ramsey2017million} model). 
Regardless of a causal structure's construction method, the use of DCGs within visualization and visual analytics can be a useful approach for both confirmatory and exploratory analysis.}

\edits{Yet despite their value, these types of causal models also have key limitations. For example, manual construction methods rely heavily on the assumptions and expertise of the person defining the model~\cite{cooper2019handbook}. Moreover, they typically result in small DCGs with relatively few variables~\cite{nordon2019building}.
Statistical DCG models, meanwhile, have been reported to have difficulty scaling to large and complex datasets without introducing significant errors and bias~\cite{lipsky_causal_2022}.}
\edits{Some approaches, such as Exploratory Factor Analysis (EFA)~\cite{fabrigar2011exploratory} have been proposed to overcome the scalability limitations of DAG-based applications, but they are only partial solutions.} 

\edits{More details about limitations are discussed in the remainder of this section, and these limitations in part motivate the proposed directions outlined in this article.  However, we emphasize that counterfactual-based approaches can lead to additional methods to support visual causal inference that work as a \emph{complement} to traditional DCG-based approaches and \emph{not} as a replacement.}

\subsection{Data Quality}
Data quality is one of the most crucial concerns that could impact the performance of statistical models.
Most datasets are not originally created for causal inference tasks and can exhibit many quality issues that raise questions about the validity of any automatically constructed graphical causal model. 
More specifically, statistical models of any kind--including graphical causal models--can be quite sensitive to noisy, incomplete, missing, invalid, confounded, or unrecorded data. Sadly, these limitations are common in real-world datasets.
Furthermore, statistical causal inference requires well-defined interventions and sufficient variation in the data to overcome confounders and identify meaningful causal effects. This makes it even harder to apply effectively to general datasets.

\subsection{Data Complexity}
The rapid development of the big data era introduces another major challenge---data complexity.
Many datasets contain a very large number and variety of variables.
For example, in data-driven healthcare applications, a single dataset can contain hundreds of thousands of unique variables~\cite{gotz_data-driven_2016} including demographics, drugs, findings from imaging, diagnoses, clinical test results, etc.
In these real-world scenarios, statistical causal models face enormous computational challenges in terms of scalability.  Calculating meaningful and comprehensive causal graphs from such a vast number of variables, interactions, and potential feedback loops is not practically achievable in most cases. 

For this reason, most applications of causal graph modeling focus on relatively small graphs with a very limited number of variables and interactions in which the computational complexity can be effectively managed. Once the problem is reduced to a small enough complexity, computational methods for mining causal graphs (or workflows that require manual specification of the causal graph) become possible. 

Unfortunately, however, this reduction of the problem also means that the resulting models are often insufficiently complex to capture real-world interactions accurately.  
Meanwhile, the ever-increasing complexity of datasets (and the exponentially increasing number of potential variable interactions that result) means that this problem will continue to grow even more difficult for traditional graphical model-based methods to overcome. 

\subsection{Direction of Causal Relationships}
In graphical causal models, detecting the presence of a causal relationship between variables is not the only goal. Causal models must also capture the direction of node links to represent the direction of causal relationships, i.e., the ``from'' nodes have a causal effect upon the ``to'' nodes for each link. This directionality adds to the complexity challenge mentioned earlier, with similar results: causal graph modeling approaches are typically limited to very small numbers of variables and potential causal relationships, and the applicability of these methods to complex real-world problems is limited unless the scope of the problem is dramatically narrowed and simplified.

\subsection{Limited Inference Levels}
Previous work has classified causal inferences along three progressive levels~\cite{pearl_causal_2016}: prediction, intervention, and counterfactual reasoning.
However, existing visual analytics systems primarily focus on the two lower levels of this framework--- prediction and intervention, while neglecting the third level of inference: counterfactual reasoning.

This limitation may be in part related to the widespread use of DAG models which do not explicitly capture reciprocal and counterfactual effects. DAGs assume unidirectional and acyclic causal relationships between two variables. However, even for DCGs which allow cycles, graphical approaches largely focus on deriving and communicating the existence and magnitude of pairwise causal relationships rather than counterfactual reasoning about what would happen under alternative conditions which is the hallmark of counterfactual reasoning.

\section{\MakeUppercase{Counterfactuals}}

Some aspects of the challenges outlined in the prior section, such as data quality, are ones that can be addressed in part with better data gathering and archiving practices. However, the remainder deal with issues of scale and complexity which render traditional graphical causal model-based approaches impractical for many real-world problems. Instead, we argue that an alternative approach which builds on the foundational concept of counterfactual reasoning can provide many key benefits in the context of supporting causal reasoning while offering a more practical solution for dealing with data complexity. 

\begin{figure*}
\centerline{\includegraphics[width=38pc]{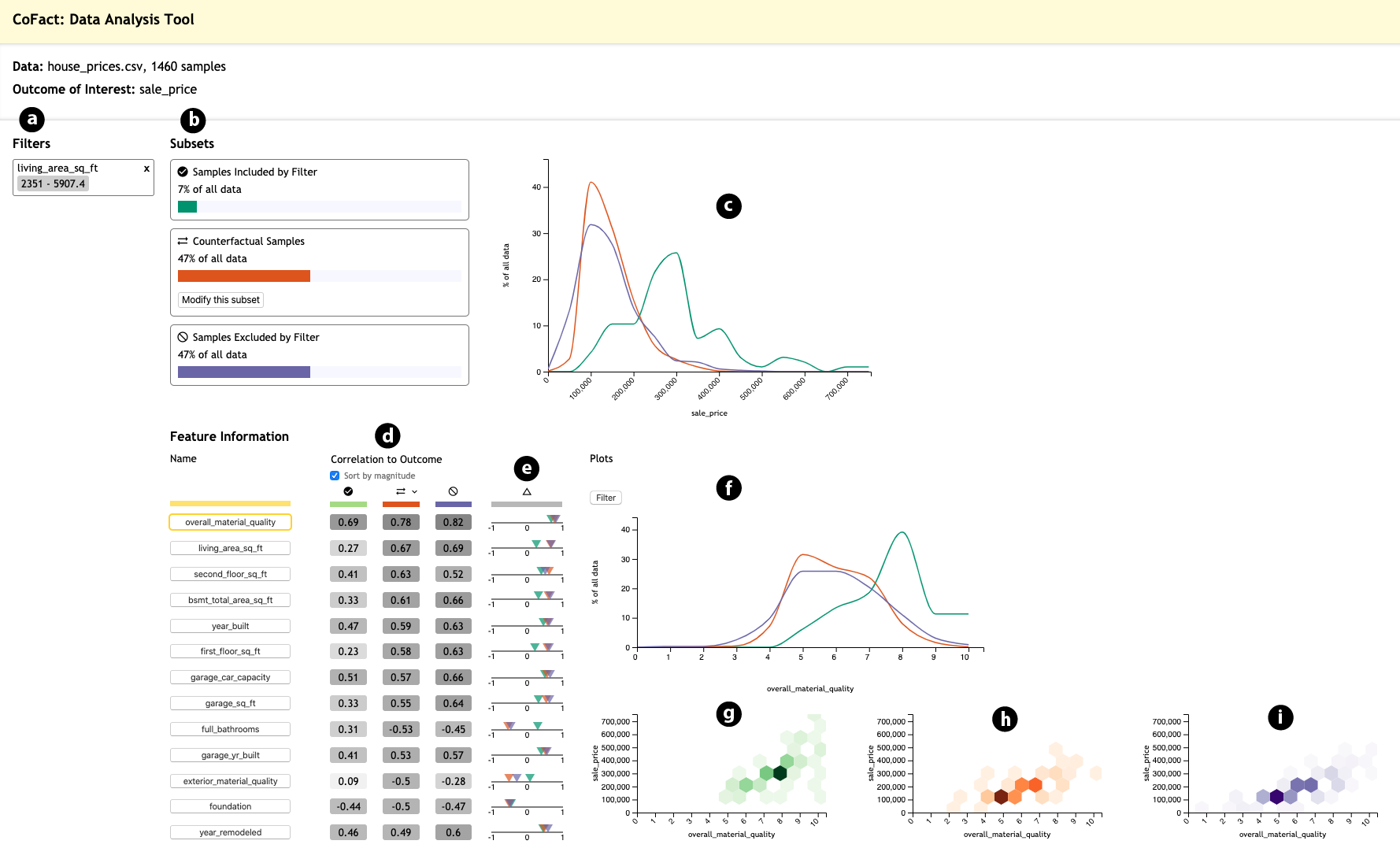}}
\caption{The CoFact visualization system~\cite{kaul_improving_2021} leverages counterfactuals to help better communicate the relationship between variables of interest. \edits{In this figure's example, the user has applied a filter constraint on square footage (a) to a multidimensional house sales dataset. In response, they are shown the resulting included, counterfactual, and excluded subsets (b), along with their corresponding distributions for a selected outcome feature of interest: house sale price (c). Additional feature-to-outcome relationships can be explored with supplementary visualizations (d–i).} The tool supports comparisons between an included subset (data points that match user-specified inclusion criteria) and a counterfactual subset containing similar data points selected from those data points that do not meet the inclusion criteria.}
\label{cofact_fig}
\end{figure*}

\subsection{What is a Counterfactual}
Counterfactuals are a core philosophical construct that underpins modern causality theory~\cite{lewis_counterfactuals_2013, hume_treatise_1978}. Counterfactual thinking posits that if $A$ causes $B$, then in an alternative, ``counterfactual'' scenario where $A$ does not occur, $B$ should not occur. Counterfactual thinking also asks us to investigate possible scenarios in which $A$ does not occur but $B$ occurs nonetheless. Such a scenario suggests that $B$ may in fact be caused by factors other than $A$. Byrne~\cite{byrne_counterfactuals_2019} adds that counterfactuals can serve an explanatory function, amplifying causal judgment. For instance,  if one could know that an alternative scenario that eliminates $A$ would not lead to $B$, it would amplify one’s judgment of a causal relationship between $A$ and $B$.  In contrast, knowing that an alternative scenario eliminates $A$ but also leads to $B$ would weaken confidence in the causal influence of $A$ on $B$.

As a concrete version of this idea, consider the example from this paper's introduction about the happiness of those active on social media vs. the happiness of those who do not use social media.  If those without social media are identical to those who do use social media in every way except for their social media usage (they have the same number of close friendships at work, the same connections to neighbors, etc.), then that group serves as a counterfactual to the social media users. This gives us confidence that in fact the only remaining difference between the populations---the degree of social media use--is causally linked to the differences in happiness.  

The caveat to this approach, of course, is that true counterfactuals must be identical in every way except for the factor being considered for causal effect, i.e. identical in every way except for their use of social media. This is possible, perhaps, in philosophical discussions of causality. However, in practice, we are typically limited to ``highly similar'' instead of identical, as many factors are unobserved. Moreover, even those factors that are accurately captured within a dataset typically exhibit some degree of variance.

Yet despite these practical limitations, counterfactual-based approaches are widely used in machine learning for tasks such as prediction (e.g., ~\cite{prosperi2020causal}), explainable AI (e.g.,~\cite{gomez_vice_2020}), and fairness (e.g.,~\cite{kusner_counterfactual_2017}). 
\edits{In contrast, counterfactuals have been less commonly used within the visualization community. Moreover, we argue that there could be opportunities to leverage this concept in new ways for visualizations to support users' improved causal reasoning.}

\subsection{Visualizing Counterfactuals}
To explore the potential benefits of leveraging counterfactual-based reasoning within the context of visualization, we recently conducted a study using an early prototype---named CoFact---that leveraged similarity-based counterfactuals~\cite{kaul_improving_2021}. This pilot project, a screenshot of which can be seen in \autoref{cofact_fig}, explored how counterfactuals could be used when filters are applied to narrow the focus of analysis during exploratory visualization. More specifically, this project proposed the definition of several important subsets when visualizing data for a given combination of filtering constraints.

First, CoFact defined the \emph{included} data subset as the portion of a dataset that meets the inclusion criteria.  Second, it defined the \emph{excluded} data subset as the portion of a dataset that did not meet the inclusion criteria. In a classic design, a visualization system might visualize both the included data and the excluded data to enable users to compare the two subgroups. For example, in our motivating social media scenario, a visualization might show happiness levels for social media users (the \emph{included} subgroup) against those who don't use social media (the \emph{excluded} subgroup). In this classic design, users might incorrectly assign a causal effect to the inclusion criteria (use of social media) to explain visualized differences in happiness.

\begin{figure*}
\centerline{\includegraphics[width=38pc]{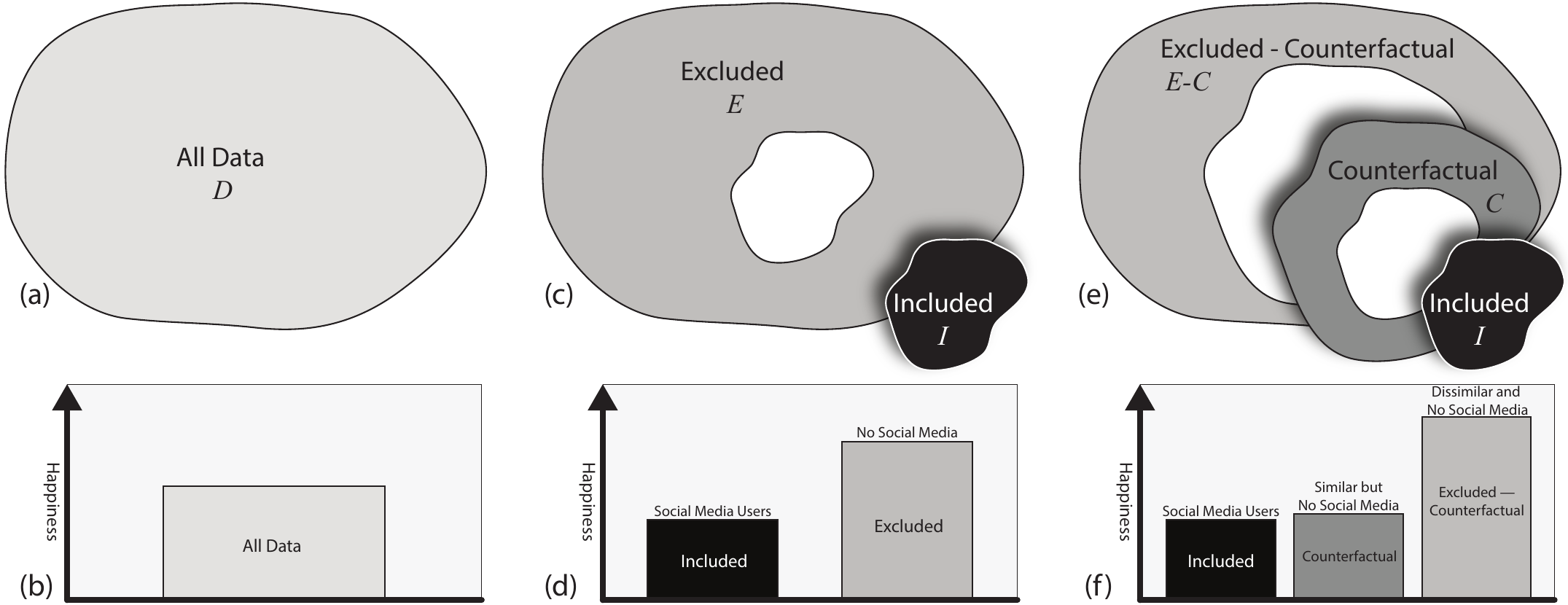}}
\caption{A counterfactual subset contains data points from the excluded set that are the most similar to those in the included set. Prior work~\cite{kaul_improving_2021} has shown that a visualization that allows users to compare the counterfactual subset against the included subset (c) supports more accurate causal inferences compared to a more traditional approach (b).}
\label{cf_set_fig}
\end{figure*}

CoFact then went beyond this classic approach to additionally define the \emph{counterfactual} subset as the portion of the \emph{excluded} subset that is most similar to the \emph{included} subset. \edits{CoFact computed a \emph{Euclidean} distance measure from each data point in the \emph{excluded} subset to all data points in the \emph{included} subset to determine a \emph{counterfactual} subset that contains the closest data points. This process is illustrated in \autoref{cf_set_fig}.} In our example scenario, the \emph{counterfactual} subset would be people from the \emph{excluded} ``do not use social media'' subset who are most similar to people from the \emph{included} ``social media users'' subset across all other dimensions in the data. In other words, the \emph{counterfactual} subset would include only the excluded people who could best serve as counterfactuals to the included subset.
\edits{Various other \emph{matching} methods have been proposed in the statistical causal inference literature, such as propensity score matching~\cite{caliendo2008some} and Mahalanobis distance matching~\cite{king_nielsen_2019}. Such methods could also be applied within CoFact to determine the counterfactual subset.}

CoFact automatically derived and visualized this \emph{counterfactual} group for comparison against the \emph{included} subset during exploratory analysis. When studying the effect of this use of counterfactual information during visualization, we found that users presented with a visualization of the \emph{counterfactual} subset were significantly more successful at identifying spurious correlations 
\edits{(i.e., correlations unlikely to be indicative of a causal relationship) that were more accurately explained by relationships to other variables within the dataset.} When using a control version of the system that did not present counterfactual data, users were significantly more likely to incorrectly assign causal effects to these same spurious relationships. A more detailed presentation of the study results and a discussion of its implications are available in the original publication~\cite{kaul_improving_2021}.

\subsection{Advantages and Limitations}

Counterfactual visualizations such as those used in CoFact offer several advantages over alternatives such as visual causal analysis using causal graphs.  First, unlike causal graph mining, counterfactuals can be scalably calculated and used with both high-volume and high-dimensional data. Second, the implementation required to compute counterfactual subsets is relatively simple and straightforward making it more transparent and explainable.  Moreover, the approach lends itself to easy integration into many existing visualization workflows without requiring that users learn about causal graphs and other advanced concepts. 

It is also important to note that this approach is not tied to any specific visualization type, but rather can be used in conjunction with a wide variety of representations including basic statistical graphics (e.g., bar charts, line charts, scatter plots) as well as more complex or bespoke visual designs.
\edits{Moreover, counterfactuals are not limited to scenarios where causal structures are computationally derived from data. They can offer significant value in contexts where causal models are predefined as well, such as in controlled experimental settings. Counterfactual analysis can be used to help confirm or refute prior causal assumptions, providing a powerful means to assess existing models.}

However, counterfactual visualizations also have some important limitations to consider. First, the entire approach requires data subsets (in order to form subsets for comparison). As a result, they are not useful when looking at overviews of entire datasets. Second, the approach depends on the identification of a ``good'' counterfactual subset which can be difficult to quantify and at times may not be present in the data. \edits{This remains an open challenge, though alternative similarity methods based on information-theoretic metrics such as entropy, or topological metrics, could be investigated as tools to help identify ``good'' subsets.} Finally, adding counterfactuals to visualizations can make the visual representation more difficult to interpret and cause analysts to work more slowly, as there is more information to process.

\section{\MakeUppercase{Future Opportunities for Visual Causal Inference}}

While our work on CoFact and the other related studies presented in this article have started to answer several interesting questions related to visual causal inference, they have also introduced several new opportunities for future research and experimentation. Informed by both our own work as well as related research from many others, we have identified several important topics that demand further attention in future research.  Advances on these topics would both help address some of the limitations we enumerated in the prior section as well as help advance our ability to create better visual analytics tools that help users draw improved causal inferences from complex data.

\subsection{Better Cognitive Models of Causal Inference with Visualization}

As noted previously, multiple studies (e.g., \cite{kaul_improving_2021, xiong_illusion_2020}) have shown that users draw causal inferences even from traditional visualization designs. \edits{In addition, a preliminary theoretical model from mathematical psychology has been applied to help understand how these causal inferences are made~\cite{kale_causal_2021}.} These studies suggest that aspects of a visualization's design can have an effect on the magnitude of the causal relationships that users perceive. \edits{However, we still lack a well-grounded high-level understanding of how human cognition forms causal inferences in complex contexts~\cite{griffiths2005structure}.
Furthermore, cognitive biases and illusions can affect the assessment of causal relationships~\cite{matute2015illusions}.}

\edits{For these reasons, a grand challenge in this area is the development of a rigorous and comprehensive model that accurately captures the way in which users cognitively approach analytical questions and draw causal conclusions.
Such a model would need to incorporate aspects of 
a visualization's design, the data that users have access to, the users' levels of expertise, cognitive states and biases, as well as other related factors.} If achieved, such a model would greatly advance our understanding of how users think about causal relationships and help guide the creation of a new generation of visual analytics tools.

\subsection{Improving Communication of Counterfactual Visual Representations}

Our initial research on exploring counterfactual visualizations, embodied in our CoFact system~\cite{kaul_improving_2021}, has demonstrated that even relatively naive approaches to incorporating counterfactuals into visualization workflows can help improve the accuracy of users' causal inferences. However, our evaluations also showed that the additional information could increase the complexity of a visual analytic tool's interface and require extra time to interpret. This can result in slower analytic performance and, potentially, confusion about what is being visualized. This is in part because the concept of counterfactuals is not necessarily familiar to many users of visual analytics software. 

Advances in our understanding of how best to communicate counterfactual information within a visualization is a research question that requires additional attention. Improvements in how we communicate counterfactual information will make counterfactual-based methods more accessible to a larger audience, and will potentially help users work more quickly while maintaining the quality of their visual causal inferences.

\subsection{Advances in Measures for Evaluating the Quality of Counterfactual Subsets}

Another open question centers on what makes a ``good'' counterfactual subset. This will require a deeper understanding of how to identify similar subsets within complex high-dimensional data. We note that this is an especially difficult problem because similarity is inherently both a task- and data-dependent question. Even for the same dataset, the ``most similar'' data points may be different depending upon which analytical question is being asked. Moreover, even if the most similar data points can be reliably identified, we must also understand what constitutes ``good enough'' to justify a given conclusion about the causal relationship between variables.

\subsection{Improving Guided Exploration}

Counterfactual-based visualizations have an additional potential benefit in the context of guided exploration.  We have already discussed how counterfactuals have the potential to help users make better causal interpretations of their data. This is accomplished by providing users with the counterfactual subset as a more appropriate comparison for the included set. We believe that this approach could potentially serve as the basis for improved techniques that help guide users toward more statistically interesting subspaces of their data for future analysis. Currently, many guidance approaches rely on correlation. However, incorporating counterfactual concepts may help researchers develop more effective guidance techniques which help users avoid spurious correlations and instead navigate toward visualizations that depict more meaningful causal relationships.

\section{CONCLUSION}

The tendency for consumers of visualizations to draw causal inferences based on non-causal relations and incomplete evidence is unavoidable. This inclination can lead to incorrect conclusions being drawn from data \edits{along} with subsequent impaired decision-making. Visualization designers who care about the accuracy of user interpretations should therefore employ methods to improve causal inferences. Although a body of work exists integrating statistical causal models, such as DAGs, into visual analytics tools to aid in causal reasoning, there are shortcomings to such approaches, especially with respect to dataset size and complexity. 

Recent work investigating the use of counterfactuals with visualization and visual analytics systems has shown promise as a practical, general-purpose method that scales well and integrates easily with common visualization workflows and visual designs. Counterfactuals can encourage users to think more deeply about a dataset and investigate relationships between variables that can help confirm or deny assumptions of causality. However, there remain several open challenges that must be solved for counterfactual approaches to reach their full potential. The visualization community should take these challenges on through new research that enables both advances in foundational theories of counterfactual visualization as well as applications that depend on more accurate and reliable visual causal inference.

\section{ACKNOWLEDGMENTS}

This work is supported in part by Award \#2211845 from the National Science Foundation.

\bibliographystyle{IEEEtran}
\bibliography{refs}

\begin{IEEEbiography}{David Borland}{\,}is an Assistant Director of Analytics and Data Science with RENCI at The University of North Carolina at Chapel Hill. He is involved in various research projects involving data visualization, visual analytics, and image analysis. Borland received his Ph.D. in Computer Science from The University of North Carolina at Chapel Hill. He is a corresponding author for this article. Contact him at borland@renci.org.
\end{IEEEbiography}

\begin{IEEEbiography}{Arran Zeyu Wang}{\,}is a Ph.D. student in Computer Science at The University of North Carolina at Chapel Hill. His research interests include visual analytics and graphical perception. Contact him at zeyuwang@cs.unc.edu.
\end{IEEEbiography}

\begin{IEEEbiography}{David Gotz}{\,}is a Professor of Information Science at The University of North Carolina at Chapel Hill. He leads the Visual Analysis and Communications Lab (VACLab) where he conducts research related to the study and development of visual methods for information analysis and communication. Gotz received his Ph.D. in Computer Science from The University of North Carolina at Chapel Hill. He is a corresponding author for this article. Contact him at gotz@.unc.edu
\end{IEEEbiography}

\end{document}